\documentclass[conference]{IEEEtran}
\IEEEoverridecommandlockouts
\usepackage{cite}
\usepackage{amsmath,amssymb,amsfonts,latexsym}
\usepackage{algorithmic}
\usepackage{graphicx}
\usepackage{subcaption}
\usepackage{textcomp}
\usepackage{xcolor}
\def\BibTeX{{\rm B\kern-.05em{\sc i\kern-.025em b}\kern-.08em
    T\kern-.1667em\lower.7ex\hbox{E}\kern-.125emX}}

\usepackage{mathtools}
\usepackage{comment}
\usepackage{hyperref}
\usepackage{booktabs} 
\usepackage{url}            

\ifCLASSINFOpdf
\else
\fi

\usepackage{amssymb}
\usepackage{amsmath}
\usepackage{titlesec}
\usepackage{url}
\usepackage{makecell}
\usepackage{xcolor}

\begin{document}
\title{Advancing Quantum Reinforcement Learning with Differentiable Quantum Architecture Search and Adaptive Non-local Observables}

\author{Junghoon Justin Park$^1$, Kyumin Lee$^2$, Anna Jung$^3$, Samuel Yen-Chi Chen$^4$,\\Huan-Hsin Tseng$^5$, Shinjae Yoo$^5$, Jiook Cha$^{1,6,7}$\\
\small $^1$Interdisciplinary Program in Artificial Intelligence, Seoul National University \\
$^2$School of Civil, Architectural Engineering and Landscape Architecture, Sungkyunkwan University \\
$^3$Department of English Language and Literature, Seoul National University \\
$^4$Wells Fargo \quad $^5$Artificial Intelligence Department, Brookhaven National Laboratory \\
$^6$Department of Brain and Cognitive Sciences, Seoul National University $^7$Department of Psychology, Seoul National University \\
\small \texttt{utopie9090@snu.ac.kr, supermini713@skku.edu, annajung227@snu.ac.kr, ycchen1989@ieee.org,} \\
\small \texttt{htseng@bnl.gov, sjyoo@bnl.gov, connectome@snu.ac.kr}
}


\maketitle

\begin{abstract}
Quantum Reinforcement Learning (QRL) using variational quantum circuits (VQCs) is constrained by fixed measurement protocols and hand-designed circuit ansätze. Adaptive Non-local Observables (ANO) and Differentiable Quantum Architecture Search (DiffQAS) address these issues independently, but to our knowledge have not been combined or compared in a single QRL framework. We present an ablation of ANO and DiffQAS within the Quantum Time-series Transformer (QTSTransformer), evaluated on a $2 \times 2$ factorial design across six environments---two classic-control tasks and four Atari 2600 games. Three findings emerge. First, the quantum agent outperforms a matched classical transformer baseline on five of six environments, with absolute scores ranging from 360.0 vs.\ 304.0 (MarioBros) to 17.2 vs.\ 2.5 (Breakout), while using approximately $12.5\times$ fewer transformer parameters on the four Atari environments (more modestly so on the two classic-control tasks). Second, ANO and DiffQAS provide complementary benefits: ANO yields the largest gain on continuous-state CartPole (347.7 vs.\ 223.2), while DiffQAS yields the largest absolute Atari gains (Breakout 17.2 vs.\ 3.0; DonkeyKong 70.0 vs.\ 37.0).
Third, combining both innovations exhibits an asymmetric interaction: ANO-dominant tasks tolerate DiffQAS, whereas DiffQAS-dominant tasks degrade when ANO is added. These results provide design guidance for QRL agents: select between ANO and DiffQAS according to whether the bottleneck is the measurement basis or the gate topology, rather than applying both indiscriminately.
\end{abstract}


%
\IEEEpeerreviewmaketitle

\section{Introduction}
Quantum Reinforcement Learning (QRL) has emerged as a promising frontier in quantum machine learning, using quantum parallelism and entanglement for potential gains in convergence and representation power \cite{Kaldari}. Variational Quantum Circuits (VQCs) have been deployed as function approximators across classic control, arcade games, robotics, and financial prediction tasks \cite{SamuelChen_2020_QRL, Skolik, Lockwood, Hohenfeld, Dutta}. However, current QRL research remains constrained by two bottlenecks: fixed measurement protocols and manually-crafted circuit architectures.

The first limitation is the measurement layer: standard VQCs use fixed local observables (e.g., single-qubit Pauli-Z), restricting output to a confined eigenvalue spectrum that fails to capture complex state-action correlations \cite{Lin_ANO}. Adaptive Non-local Observables (ANO) address this by treating the Hermitian observable as a learnable parameter, letting the model expand its functional space and capture non-local correlations without added circuit depth \cite{Lin_ANO}, adapting its measurement basis to the reward landscape \cite{Lin_ANO_QRL}.

The second hurdle is VQC architectural design: most QRL agents rely on hand-crafted ans\"atze requiring domain expertise and lacking flexibility across environments \cite{SamuelChen_2024}. Differentiable Quantum Architecture Search (DiffQAS) relaxes the discrete search space into a continuous, gradient-optimizable domain \cite{Zhang_2022_DiffQAS}, in principle allowing task-specific architecture selection and reducing the manual trial-and-error burden \cite{Sun, SamuelChen_2024}.

Despite the individual successes of ANO and DiffQAS, no study to date has combined or systematically compared these two paradigms, leaving open whether adaptive measurements, automated architectures, or their combination is optimal for high-dimensional sequential decision problems.
We address this gap by proposing a QRL framework based on the Quantum Time-series Transformer (QTSTransformer) \cite{Park_QTSTransformer}, which integrates both DiffQAS and ANO. By evaluating our model on a spectrum of environments—ranging from simple OpenAI Gymnasium tasks to complex, pixel-based Atari 2600 games—this work provides a systematic analysis of how learnable measurements and automated architecture search collectively influence the performance and scalability of QRL agents.

\section{Background}
\subsection{Quantum Reinforcement Learning (QRL)}
Reinforcement learning models sequential decision-making as a Markov Decision Process $\langle S, A, R, P, \gamma \rangle$; the agent learns a policy $\pi$ maximizing the expected \emph{return} $G_t = \sum_{k=0}^\infty \gamma^k r_{t+k}$ (not the immediate reward $r_t$) by approximating a value function $Q(s,a)$ \cite{Sutton, Mnih}.

Quantum Deep Quality-Networks (DQN) replaces the classical DQN with a VQC-based function approximator \cite{SamuelChen_QRL}: a state $s$ is encoded via $U(s)$ and processed by a trainable ansatz $W(\theta)$. Rather than the textbook single-observable-per-action formulation, most practical QRL implementations---including ours---measure a fixed set of $N_O$ observables independent of $|\mathcal{A}|$ and feed the expectation values into a classical fully-connected (FC) head layer $f_\text{head}: \mathbb{R}^{N_O} \to \mathbb{R}^{|\mathcal{A}|}$:
\begin{align}
\label{eq:q_pipeline}
    Q(s, \cdot; \theta, \phi) \;&=\; f_\text{head}\!\left(\bigl\{\langle 0 | U(s)^\dagger W(\theta)^\dagger \hat{O}_i W(\theta) U(s) | 0 \rangle\bigr\}_{i=1}^{N_O}\right) \nonumber \\ &\in \mathbb{R}^{|\mathcal{A}|},
\end{align}
where $\{\hat{O}_i\}_{i=1}^{N_O}$ are fixed Pauli operators or learnable ANO Hermitians; this accommodates $|\mathcal{A}|$ varying across environments (2 for CartPole, up to 18 for DonkeyKong/MarioBros) at a fixed circuit width of 8 qubits ($N_O=24$ baseline, $N_O=8$ ANO; mapping detailed below).

To stabilize training we use experience replay and the \emph{Double DQN} target \cite{Hasselt}, which decouples action selection (online network $\theta$) from value evaluation (target network $\theta^-$) and reduces the upward bias of the standard $\max_{a'}$ operator.

\subsection{Quantum Time-series Transformer (QTSTransformer)}
The Quantum Time-series Transformer (QTSTransformer) is a hybrid sequence-processing architecture whose idealized quantum subroutine for the time-mixing operation has cost $O(\text{polylog}(T))$ per query in sequence length $T$, compared to $O(T^2)$ for explicit classical self-attention~\cite{Park_QTSTransformer}. This asymptotic statement applies only to the abstract subroutine; in this paper the entire pipeline runs on a classical state-vector simulator, and end-to-end cost is dominated by convolutional neural network (CNN) preprocessing, the FC head, replay, and simulator overhead. We therefore use the QTSTransformer as an architectural and inductive-bias choice, not as a wall-clock speedup.
The architecture has four stages:

A sequence of latent feature vectors $X = \{x_0, \dots, x_{T-1}\}$ is mapped to rotation angles that parameterize unique unitaries $U_j(\theta_j)$, encoding temporal information directly into a quantum gate sequence. Time-mixing is motivated by the linear-combination-of-unitaries (LCU) primitive \cite{Childs}: a control register in superposition directs the application of sequence unitaries $\{U_0, \dots, U_{T-1}\}$ on a target register, yielding $\hat{M} = \sum_{j=0}^{T-1} e^{i\gamma_j} |a_j|^2 \, U_j$, where $|a_j|^2$ is the (unnormalized) attention weight for time step $j$ (Born-rule weight from the control register).

Non-linearity is motivated by the quantum singular value transformation (QSVT) framework \cite{Gilyen}, expressed as a degree-$d$ polynomial $P_c(\hat{M}) = c_d \hat{M}^d + c_{d-1} \hat{M}^{d-1} + \dots + c_1 \hat{M} + c_0 I$ acting on the singular values of an appropriately block-encoded $\hat{M}$.
We evaluate $P_c(\hat{M})|0\rangle$ by accumulating $\sum_k c_k \hat{M}^k |0\rangle$ with learnable $\{c_k\}$ (degree $d=2$ in all runs).
This state is then renormalized and passed through one further single-layer ansatz---the quantum feed-forward block---whose four gate slots are included in the architecture-parameter count of Table~\ref{tab:params}. The resulting state is measured to produce $N_O$ expectation values $\langle\hat{O}_1\rangle,\dots,\langle\hat{O}_{N_O}\rangle$, with $\hat{O}_i$ either fixed single-qubit Paulis ($N_O=24=3\times 8$) or learnable $k=2$-local Hermitians on 8 sliding windows ($N_O=8$), processed by a compact FC head---Linear($N_O$, 128) $\to$ ReLU $\to$ Linear(128, $|\mathcal{A}|$) on Atari, Linear($N_O$, 64) $\to$ ReLU $\to$ Linear(64, $|\mathcal{A}|$) on classic control---to produce the $|\mathcal{A}|$-dimensional Q-value vector, realizing Eq.~\eqref{eq:q_pipeline} uniformly across all environments.


\subsection{Adaptive Non-local Observables (ANO)}
\label{sec:ano_bg}
As noted above, fixed local measurements confine VQC output to $[-1, 1]$, limiting expressivity where $Q$-values span broader ranges. ANO addresses this by treating the Hermitian measurement operator as a learnable parameter \cite{Lin_ANO, Lin_ANO_QRL, SamuelChen_LearningtoMeasure}.

Inspired by the Heisenberg picture, ANO makes the readout observable $H$ trainable rather than fixed---as opposed to standard fixed-Pauli VQCs, restricted to observables in the spectral equivalence class of the chosen Pauli basis. While the fully general case spans the full $4^n$-dimensional Hermitian space \cite{Lin_ANO}, our implementation parameterizes Hermitians on $k$-qubit subsystems only, so the agent navigates the space of $k$-local Hermitians on the chosen sliding windows (we use $k=2$).\label{sec:ano_param} For a $k$-qubit system, let $K = 2^k$; an adaptive $k$-local observable $H(\phi)$ is Hermitian by design, built from a lower-triangular matrix $h$ with three sets of learnable parameters indexed by $i > j$---real strictly-lower entries $\{a_{ij}\}$, imaginary strictly-lower entries $\{b_{ij}\}$ (each $K(K{-}1)/2$ parameters), and real diagonal entries $\{d_{ii}\}$ ($K$ parameters)---giving $H(\phi) = h + h^\dagger$, where $\phi = (a_{ij}, b_{ij}, d_{ii})$ represents $K^2$ trainable real parameters in total: diagonal entries $2d_{ii}$ and Hermitian-conjugate off-diagonal pairs $a_{ij} \pm i b_{ij}$, guaranteeing $H(\phi) = H^\dagger(\phi)$ and real-valued expectation values.

By the Rayleigh quotient, the output of an ANO-VQC is bounded by the spectrum of $H(\phi): \lambda_{\min} \le \langle \psi | H(\phi) | \psi \rangle \le \lambda_{\max}$. Strictly, this argument only shows that the achievable output \emph{range} becomes learnable; it does not by itself imply improved nonlocal-correlation modeling. Capturing 2-local correlations is plausible since each $4 \times 4$ Hermitian on a 2-qubit subsystem is generally not separable into single-qubit Paulis \cite{Lin_ANO}, but isolating range from correlation effects would require a fixed-spectrum 2-local control that we do not include here. In a QRL setting, the agent jointly optimizes the variational parameters $\theta$ and the observable parameters $\phi$ by minimizing the same Double DQN loss, now taken jointly over $(\theta, \phi)$ with target-network copies $(\theta^-, \phi^-)$; since $H(\phi) = h + h^\dagger$ is differentiable with respect to all learnable entries of $h$, the gradients $\partial \mathcal{L} / \partial \phi$ are computed via automatic differentiation through the quantum circuit simulator \cite{SamuelChen_LearningtoMeasure}.

\subsection{Differentiable Quantum Architecture Search (DiffQAS)}
As noted above, DiffQAS relaxes the discrete choice of quantum gates into a continuous, end-to-end optimizable formulation \cite{Zhang_2022_DiffQAS, Sun, SamuelChen_2024, SamuelChen_DiffQAS_QLSTM}. The general framework \cite{Zhang_2022_DiffQAS, SamuelChen_DiffQAS_QLSTM} assigns each candidate gate a learnable structural weight, relaxed via softmax into a weighted ensemble over candidates, and discretizes to the highest-weight candidate upon convergence.

\paragraph{Euler Angle Parameterization}
\label{sec:euler}
An alternative to the weighted ensemble is to parameterize each gate slot via continuous Euler angles, allowing smooth interpolation between canonical gates within a single circuit \cite{Sun}. Each single-qubit gate is replaced by $\text{Rot}(\varphi, \alpha, \omega) = R_Z(\omega)\, R_Y(\alpha)\, R_Z(\varphi)$,
where $\alpha$ is the variational angle (a single per-gate scalar, distinct from the global parameter set $\theta$ used for the rest of the network) and $(\varphi, \omega)$ are learnable architecture parameters. With $\alpha$ free, the Euler identity $R_X(\alpha) = R_Z(-\pi/2)\, R_Y(\alpha)\, R_Z(\pi/2)$ implies
\begin{equation}
\label{eq:rot_refs}
    R_Y(\alpha): (\varphi, \omega) = (0, 0), \quad R_X(\alpha): (\varphi, \omega) = (\tfrac{\pi}{2}, -\tfrac{\pi}{2}).
\end{equation}
$R_Z(\alpha)$, by contrast, \emph{cannot} be realized for arbitrary $\alpha$ by any fixed $(\varphi, \omega)$, since $R_Z(\omega) R_Y(\alpha) R_Z(\varphi)$ reduces to a $Z$-axis rotation only when $\alpha$ is an integer multiple of $2\pi$, in which case it equals $\pm R_Z(\varphi + \omega)$ and is independent of $\alpha$. The rotations realizable by the continuous parameterization are therefore $\{R_Y, R_X\}$ (and $\{CR_Y, CR_X\}$ for controlled slots).

Our implementation places the three discretization references at $(\varphi, \omega) \in \{(0, 0),\, (-\pi/2,\, \pi/2),\, (\pi/2,\, -\pi/2)\}$, labeled $R_Y$, $R_X$, and $R_Z$. Two implementation-level caveats---a sign absorbed by the learnable $\alpha$ at the $R_X$ point, and the $R_Z$ point actually realizing $R_X(\alpha)$ during Phase~1 (the search phase; Eq.~\eqref{eq:rot_refs}) before Phase~2 (the production phase) substitutes the genuine $R_Z(\alpha)$---did not affect the production-phase circuits, since all six environments converged to the $(0,0)$ initialization (a finding we interpret with caution below). Architecture parameters start at $(0,0)$ ($R_Y/CR_Y$), are jointly optimized with $\alpha$ during the search phase, and are discretized to the nearest canonical reference by Euclidean distance at the end.

\section{Experiment}
We evaluate the QTSTransformer agent on classic-control tasks \cite{OpenAIGym} (for precise ANO/DiffQAS ablations) and high-dimensional Atari 2600 games \cite{Atari2600}.
Episode counts are not directly comparable because episode \emph{length} differs sharply: despite a $6\times$ gap in episode count, CartPole ($\sim$500 episodes $\approx$ 65{,}000 steps) and FrozenLake (3{,}000 episodes $\approx$ 60{,}000--78{,}000 steps) see comparable environment interactions. Atari quantum runs use $\sim$0.85--0.95M steps each versus $\sim$1.6--3.1M for the classical baselines (trained to 10{,}000 episodes). Training ends where the plateau/decline stopping rule or the 48-hour wall-clock limit triggered, not at an arbitrary cutoff.

\subsection{Classic Control Environments}
Classic control tasks provide a controlled regime for fundamental learning efficiency, with low-dimensional state vectors encoded directly into quantum circuit parameters.

\paragraph{CartPole-v1}
In CartPole-v1 \cite{OpenAIGym}, the agent balances a pole on a cart via horizontal force; the episode ends if the pole angle exceeds $\pm 12^\circ$, the cart moves more than 2.4 units from center, or 500 timesteps elapse. The 4-dimensional observation (cart position, velocity, pole angle, angular velocity) pairs with a discrete two-action space (push left/right) and a $+1$ reward per timestep upright (max return 500). A sliding window of 4 states forms a $(4, 4)$ tensor, projected via a linear layer to a 64-dimensional feature space and scaled via sigmoid to $[0, 1]$ to parameterize the rotation gates.

\paragraph{FrozenLake-v1}
FrozenLake-v1 \cite{OpenAIGym} requires navigating a $4 \times 4$ grid from a starting tile to a goal tile under stochastic ``slippery'' dynamics (only a $1/3$ chance of moving in the intended direction). The discrete state $\{0, \dots, 15\}$ is one-hot encoded (16-dimensional); the action space has 4 discrete movements. Rewards are sparse ($+1$ only on reaching the goal, 0 otherwise). As with CartPole, a sliding window of 4 timesteps forms a $(4, 16)$ tensor, mapped via a linear layer to 64 rotation parameters scaled to $[0, 1]$ via a sigmoid.

\subsection{Atari2600 Environments}
We test high-dimensional feature processing and long-range temporal modeling on four Atari 2600 games \cite{Atari2600}.

\paragraph{Common Preprocessing and Quantum Interface}
%
Atari observations are preprocessed to reduce dimensionality while preserving motion cues: actions are repeated across 4 frames (frame skipping), raw RGB frames ($210 \times 160 \times 3$) are converted to grayscale and resized to $84 \times 84$, and pixel values are normalized to $[0, 1]$ and stacked across the 4 most recent timesteps to form a $(4, 84, 84)$ input tensor.

A DQN-style CNN \cite{Mnih} extracts 128-dimensional features per timestep; standard sinusoidal positional encoding \cite{Vaswani} represents temporal order within the transformer. The resulting vectors are projected to the rotation parameter space and scaled to $[-\pi, \pi]$ via a centered sigmoid: $\theta = (\sigma(x) - 0.5) \cdot 2\pi$.

\paragraph{Benchmark Games}
\textbf{SpaceInvaders (ALE/SpaceInvaders-v5)}: a 6-action shoot-'em-up (NOOP, FIRE, LEFT, RIGHT, LEFTFIRE, RIGHTFIRE) with score-based rewards. \textbf{Breakout (ALE/Breakout-v5)}: a 4-action ball-and-paddle game (NOOP, FIRE, LEFT, RIGHT); rewards scale with brick row height, and the game accelerates as bricks clear. \textbf{DonkeyKong (ALE/DonkeyKong-v5)}: an 18-action platforming task with rewards from item collection, obstacle avoidance, and stage-completion bonuses. \textbf{MarioBros (ALE/MarioBros-v5)}: an 18-action platforming task based on the 1983 arcade game, with rewards for defeating enemies and coin collection.

\subsection{QRL Agent Design and RL Techniques}
The agent is a hybrid classical-quantum DQN with the QTSTransformer-based readout over sequences of four states (or four CNN-encoded frames for Atari). We use \emph{Double DQN}~\cite{Hasselt} and uniform experience replay. All models are trained until the 100-episode moving-average return plateaus or begins to decline.

\subsection{ANO and DiffQAS Ablation Methodology}
\label{sec:ablation_method}
We conducted a $2\times2$ factorial ablation study to isolate the contributions of our two primary quantum innovations:

\paragraph{ANO}
The ANO condition replaces the baseline's fixed Pauli readout with eight learnable $k=2$-local Hermitian observables. Each $H = h + h^\dagger$ is a $4 \times 4$ Hermitian with 16 learnable parameters (6 real off-diagonal, 6 imaginary off-diagonal, 4 diagonal). The observables are placed in circular sliding windows across 8 qubits, allowing the agent to adapt its measurement basis to the reward landscape.

\paragraph{DiffQAS}
DiffQAS automates the selection of gate types in the variational circuit using the Euler angle parameterization described earlier. Each ansatz layer contains four gate slots---two single-qubit slots searching over $\{R_Y, R_X, R_Z\}$ and two controlled slots searching over $\{CR_Y, CR_X, CR_Z\}$ (though $R_Z$/$CR_Z$ are reachable only as discretization targets, not during the continuous search itself)---parameterized via $\text{Rot}(\varphi, \alpha, \omega)$ and $\text{CRot}(\varphi, \alpha, \omega)$, respectively, where $\alpha$ is the variational rotation angle. Training proceeds in two phases: a Search Phase in which variational parameters and architecture parameters $(\varphi, \omega)$ are jointly optimized, followed by a Production Phase in which each $(\varphi, \omega)$ is discretized to the nearest canonical gate by Euclidean distance and frozen for the remainder of training.

\begin{figure*}[!t]
\centering
\includegraphics[width=0.95\textwidth]{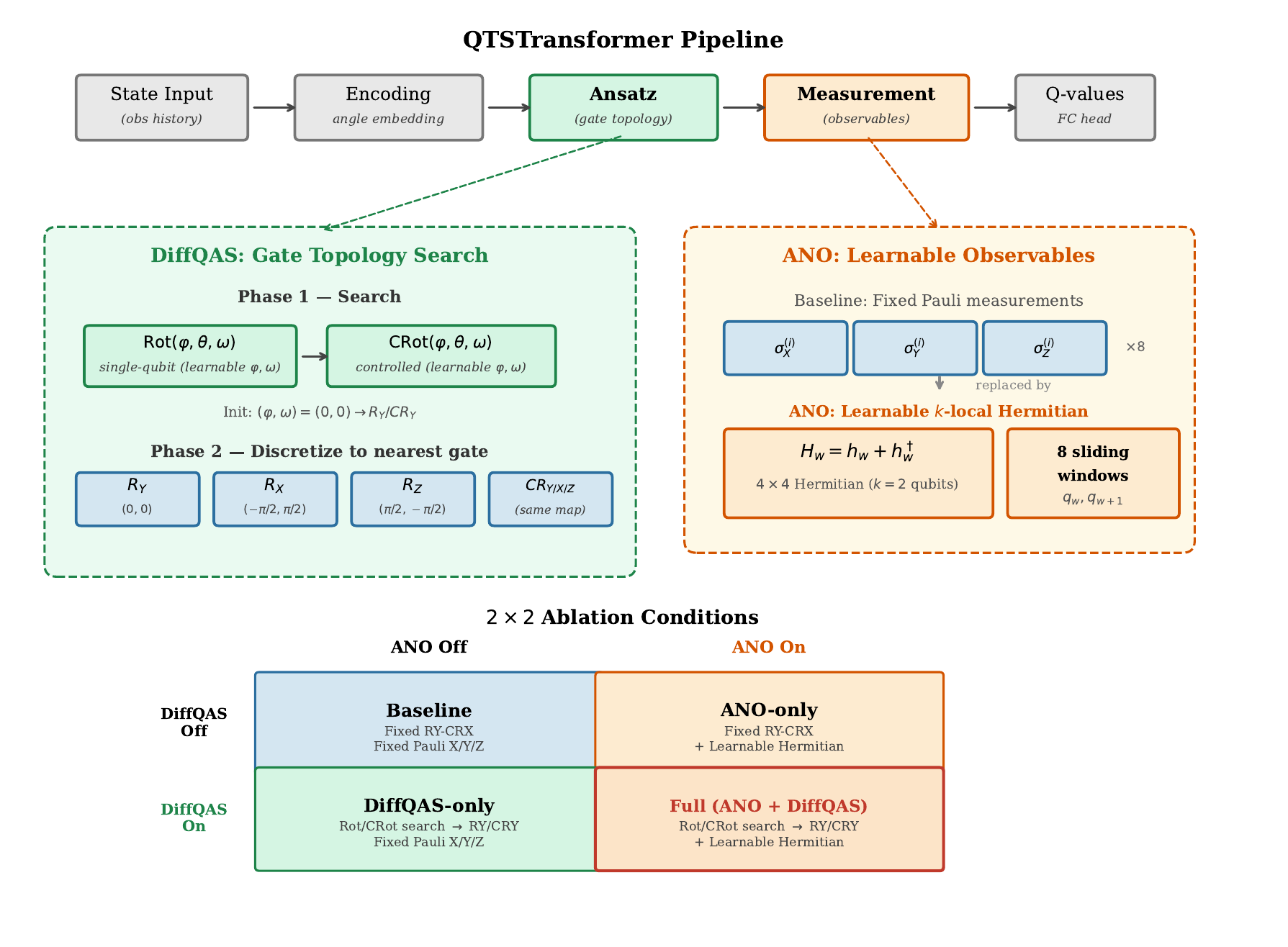}
\caption{Overview of the $2 \times 2$ ablation design. \textbf{Top}: The QTSTransformer pipeline, with the two variable components highlighted---\textcolor[HTML]{1e8449}{\textbf{Ansatz}} (gate topology, green) and \textcolor[HTML]{d35400}{\textbf{Measurement}} (observables, orange). \textbf{Middle-left}: DiffQAS's gate search. \textbf{Middle-right}: ANO's learnable observables. \textbf{Bottom}: The four ablation conditions from the factorial combination of DiffQAS and ANO.}
\label{fig:ablation_diagram}
\end{figure*}

\paragraph{Full and Baseline Models}
The Full model optimizes VQC, ANO Hermitian, and DiffQAS architecture parameters jointly. The Baseline model uses a fixed RY-CRX circuit with fixed PauliX/Y/Z measurements on all qubits (24 observables for 8 qubits). A Classical model replaces the quantum core with a standard \textit{nn.TransformerEncoder} for benchmarking; Figure~\ref{fig:ablation_diagram} overviews the pipeline and how the four ablation conditions differ in ansatz and measurement.

This Classical model is a \emph{matched-protocol baseline, not a parameter-matched or separately tuned classical RL agent}: it shares the same CNN feature extractor, replay buffer, $\epsilon$-greedy schedule, and Double DQN loop as the quantum conditions, so only the temporal module (VQC vs.\ \textit{nn.TransformerEncoder}) changes. The low classical CartPole score (132.6) is thus the ceiling of this matched-protocol transformer head under our budget, not a statement about classical RL on CartPole---a standard, tuned classical DQN or PPO solves CartPole-v1 to near-maximal return; we simply did not run one. Our claim is that the quantum core outperforms a matched classical core under the same training protocol and budget, with far fewer parameters, not that it beats the best achievable classical agent.

All conditions use 8 qubits, 2 ansatz layers, and QSVT degree 2, optimized with Adam (batch 64/32 for classic control/Atari); VQC/ANO/DiffQAS learning rates are 0.001/0.01/0.001 (classic control) and 0.00025/0.01/0.001 (Atari), with a DiffQAS search phase of 50/100 episodes, discount factor $\gamma=0.99$, $\epsilon$ schedule 1.0$\to$0.01 (decay 0.995) / 1.0$\to$0.1 (decay 0.9999), replay memory 10,000/100,000, and target-network sync every 500/1,000 steps.

\paragraph{Computational Infrastructure}
We used PyTorch (v2.5.0) and PennyLane \cite{pennylane} (v0.42.3) to implement our RL agents. All experiments ran on the Perlmutter supercomputer (NERSC, Lawrence Berkeley National Laboratory) on NVIDIA A100-80GB GPU nodes.

\section{Results}
\label{sec:results}

We evaluate the QTSTransformer agent across six reinforcement learning environments spanning two complexity regimes: classic control (CartPole-v1, FrozenLake-v1) and Atari games (SpaceInvaders, Breakout, DonkeyKong, MarioBros). For each environment, we compare four conditions in a $2 \times 2$ factorial ablation design---Baseline, ANO-only, DiffQAS-only, Full (ANO + DiffQAS)---plus a matched classical transformer baseline. All quantum conditions use 8 qubits, 2 ansatz layers, and QSVT degree 2. Performance is measured by the best 100-episode rolling-average return (BestAvgR), with Atari scores computed after the first 500 episodes to exclude exploration artifacts.

\subsection{Quantum vs.\ Classical Performance}
\label{sec:results_qvc}

Table~\ref{tab:main_results} presents the complete results across all six environments and five experimental conditions (bold/underline legend in the caption).

\begin{table*}[!t]
\centering
\caption{Performance comparison across all environments and ablation conditions. Values report the best 100-episode rolling-average return (BestAvgR); for FrozenLake, values indicate the success rate. Retention~\%: end-of-training value as a percentage of the bolded (best-quantum) condition's own peak (FrozenLake excluded --- not its story). \textbf{Bold}: best quantum condition per environment. \underline{Underline}: overall best per environment. $|\mathcal{A}|$: action space size. Each value is point estimate from a single training run.}
\label{tab:main_results}
\renewcommand{\arraystretch}{0.9}
\begin{tabular}{llccccccc}
\toprule
& & \multicolumn{4}{c}{\textit{Quantum Conditions}} & \textit{Classical} & \textit{Best Q vs.\ C} & \textit{Best-Q} \\
\cmidrule(lr){3-6} \cmidrule(lr){7-7}
Environment & $|\mathcal{A}|$ & Baseline & ANO-only & DiffQAS-only & Full & Classical & Advantage & Retention \% \\
\midrule
\multicolumn{9}{l}{\textit{Classic Control}} \\
CartPole-v1 & 2 & 223.2 & 347.7 & 24.8 & \underline{\textbf{359.9}} & 132.6 & $+171\%$ & 52.1 \\
FrozenLake-v1 & 4 & \textbf{56\%} & 49\% & \textbf{56\%} & 51\% & \underline{76\%} & $-26\%$ & --- \\
\midrule
\multicolumn{9}{l}{\textit{Atari}} \\
SpaceInvaders & 6 & 339.2 & 347.0 & 311.8 & \underline{\textbf{347.9}} & 290.8 & $+19.6\%$ & 80.9 \\
Breakout & 4 & 3.0 & 2.7 & \underline{\textbf{17.2}} & 11.1 & 2.5 & $+588\%$ & 92.9 \\
DonkeyKong & 18 & 37.0 & 43.0 & \underline{\textbf{70.0}} & 25.0 & 38.0 & $+84\%$ & 80.0 \\
MarioBros & 18 & \underline{\textbf{360.0}} & 296.0 & 344.0 & 336.0 & 304.0 & $+18\%$ & 40.0 \\
\bottomrule
\end{tabular}
\end{table*}

The quantum agent outperforms the matched classical transformer on five of six environments (Table~\ref{tab:main_results}). The percentage gains over classical are $+588\%$ for Breakout and $+84\%$ for DonkeyKong; both are computed against small classical baselines and should be read alongside the absolute scores in the table.
Because BestAvgR reports an achieved peak rather than a retained value, we also report each best-quantum condition's end-of-training retention (Table~\ref{tab:main_results}, last column): Breakout, DonkeyKong, and SpaceInvaders peaks are durable (80--93\% retained), but CartPole and especially MarioBros show substantial forgetting (52.1\% and 40.0\% retained, respectively). DonkeyKong and MarioBros---the two 18-action games---also show the highest late-training variance (coefficient of variation 1.34--2.53 and 2.55--3.82 across the four quantum conditions, respectively, versus 0.22--1.18 for the other four environments), consistent with a genuine post-exploration decline in both games rather than a stable plateau.
The quantum model also uses substantially fewer parameters throughout.

The sole exception is FrozenLake, where the classical transformer achieves 76\% success vs.\ best quantum 56\%. FrozenLake is the only environment with discrete observations (16-dimensional one-hot) and inherent stochasticity (slip probability $2/3$). The classical transformer's larger parameter budget ($\sim$2K vs.\ $\sim$1.2K) is well-suited to such tabular-like stochastic mappings, whereas the quantum circuit's rotation-angle encoding aligns more naturally with continuous states.

The episode at which best-quantum performance peaks varies widely by game (DonkeyKong $\sim$1{,}200 and MarioBros $\sim$760 to SpaceInvaders $\sim$4{,}900 and Breakout $\sim$7{,}100; full spread $\sim$565--8{,}076 across the sixteen Atari quantum configurations), yet the classical transformer, trained for up to 10{,}000 episodes, surpasses the quantum peak on no Atari game. We suspect this owes something to the VQC's Hilbert-space structure acting as a useful inductive bias~\cite{schuld2021effect,abbas2021power}, but the quantum model also has far fewer parameters, and a smaller model is often just easier to optimize; separating these factors would need a parameter-matched classical control, which we did not build here.

\subsection{Effect of DiffQAS and ANO}
\label{sec:results_ablation}


\paragraph{DiffQAS: Circuit Architecture Search.}
DiffQAS shows highly task-dependent effects (Table~\ref{tab:main_results}): on Breakout, DiffQAS-only shows a monotonic climb that separates from other conditions early in training; on DonkeyKong, it achieves the largest absolute Atari gain in the study. In both cases the search discretizes to RY/CRY rather than the baseline's RY/CRX; the evidence below indicates this reflects the fixed CRY gate rather than an adaptively discovered structure.

Across all six environments the search discretized to RY/CRY. On CartPole this topology is detrimental (24.8 vs.\ baseline 223.2, near random); on SpaceInvaders, FrozenLake, and MarioBros it gives no measurable improvement. DiffQAS therefore does not universally improve performance: it converges to a fixed alternative topology that benefits tasks for which the default RY-CRX is mismatched, and is harmful when the default is already well-suited.

This convergence to RY/CRY---the search's own initialization---held tightly: each gate slot's architecture parameters moved, on average, only 1.6\% of the distance to the nearest alternative gate reference (over all 144 slots of the DiffQAS-only and Full conditions across the six environments; exactly 0\% for Breakout). Because every peak is reached in the discretized production phase, the DiffQAS-only advantage over the RY-CRX baseline reduces to a fixed RY/CRY vs.\ RY-CRX comparison under identical readout---it reflects the CRY entangling gate rather than an adaptively discovered structure, so the outcome is governed by search-space design and initialization, not task-specific discovery. Two caveats bound this: we log only the converged parameters, not the search trajectory (a small endpoint distance is consistent with either never moving or moving and returning---certain only for Breakout at $0.0000$), and we ran no from-scratch fixed-RY/CRY control (so the search machinery need not be inert, only that it discovered nothing task-specific here).

\paragraph{ANO: Learnable Observables.}
ANO shows a complementary pattern (Table~\ref{tab:main_results}): it gives the largest gain on CartPole, and on SpaceInvaders a modest peak gain accompanies a $+11\%$ improvement in last-100-episode average, indicating better late-training stability. A similar end-of-training pattern appears on MarioBros ($+61\%$ in the last-100-episode window despite a lower peak).

ANO does not help on Breakout (2.7 vs.\ 3.0), where DiffQAS-only's large gain (17.2 vs.\ 3.0 baseline) instead points to the gate topology as the bottleneck. On FrozenLake, neither ANO-only (49\%) nor DiffQAS-only (56\%) beats the baseline (56\%); as discussed above, its discrete, stochastic observations more likely suit neither quantum intervention rather than isolating a specific measurement or topology limitation.

Together, ANO and DiffQAS act on complementary parts of the circuit: ANO optimizes \emph{what is measured}, benefiting continuous-state tasks with suboptimal default Pauli measurements; DiffQAS optimizes \emph{how entanglement is generated}, benefiting tasks whose reward landscape rewards a different connectivity.

\paragraph{Combined Model (Full = ANO + DiffQAS).}
Combining ANO and DiffQAS exhibits an \emph{asymmetric interaction} that depends on which innovation dominates in isolation.

When ANO is the dominant innovation, Full matches or slightly exceeds ANO-only (Table~\ref{tab:main_results}): on CartPole, Full is the highest quantum score even though DiffQAS-only alone is near-random; on SpaceInvaders, Full is essentially tied with ANO-only. The parametric Rot/CRot gates active during Phase~1 plausibly supply additional gradient pathways that aid ANO's measurement optimization.

When DiffQAS is the dominant innovation, Full degrades. On Breakout, Full (11.1) falls below DiffQAS-only (17.2); on DonkeyKong, Full (25.0) falls below both DiffQAS-only (70.0) and the baseline (37.0). Adding ANO's 128 Hermitian parameters and separate optimizer appears to interfere with the converged CRY topology in these reward landscapes.

When neither innovation helps in isolation (MarioBros, FrozenLake), Full is at or below the baseline (336.0 vs.\ 360.0; 51\% vs.\ 56\%).

The asymmetry has a practical reading: ANO adjusts a downstream readout largely decoupled from the circuit, so it is robust to DiffQAS, whereas DiffQAS modifies the circuit itself and is more vulnerable to interference from ANO's concurrently optimized degrees of freedom.

\subsection{Parameter Efficiency}
\label{sec:results_params}

Table~\ref{tab:params} compares parameter counts across architectures.

\begin{table}[!t]
\centering
\caption{Parameter counts for the quantum and classical transformer architectures; the quantum model uses $12.5\times$ fewer transformer-level parameters than the classical model on Atari. Transformer/VQC rows include $+128$ ANO and $+24$ architecture parameters (12 gate slots $\times$ two Euler angles: four slots in each of the two timestep ansatz layers and four in the quantum feed-forward block).}
\label{tab:params}
\renewcommand{\arraystretch}{0.9}
\begin{tabular}{lcc}
\toprule
Component & Quantum & Classical \\
\midrule
\multicolumn{3}{l}{\textit{Atari}} \\
CNN feature extractor & 1,946K & 1,946K \\
Transformer / VQC & $\sim$12K & $\sim$150K \\
\midrule
\multicolumn{3}{l}{\textit{Classic Control}} \\
Feature projection & $<$1K & $<$1K \\
Transformer / VQC & $\sim$1.2K & $\sim$2K \\
\midrule
Environments won (of 6) & \textbf{5} & 1 \\
\bottomrule
\end{tabular}

\end{table}

On Atari, the quantum model shares its CNN feature extractor with the classical \textit{nn.TransformerEncoder} ($d_\text{model}=128$, 4 heads, 1 layer, $d_\text{ff}=256$) (Table~\ref{tab:params}), isolating its $12.5\times$ parameter reduction to the temporal processing module; despite this reduction, it outperforms the classical baseline on all four Atari games.

One candidate explanation is the quantum circuit's inductive bias: an 8-qubit circuit operates in a $2^8 = 256$-dimensional Hilbert space, so the state space it acts on grows exponentially with qubit count even though the parameter count grows linearly; we offer this as a hypothesis rather than a demonstrated mechanism, since our experiments do not isolate it. The QSVT polynomial mixing supplies a temporal inductive bias different from self-attention. Model size and architectural differences remain confounded with this explanation in our design, so we read the quantum-versus-classical result as a parameter-efficiency comparison rather than as evidence of a quantum inductive-bias advantage, and we claim no asymptotic or wall-clock speedup.


\section{Conclusion}
\label{sec:conclusion}

We presented an ablation of ANO and DiffQAS within the QTSTransformer across six RL environments. Three findings stood out: the QTSTransformer outperformed a matched classical transformer on 5 of 6 environments while using substantially fewer parameters; ANO and DiffQAS provided complementary gains, with ANO strongest on continuous-state tasks and DiffQAS strongest on Atari;
and combining them showed an asymmetric interaction, with ANO-dominant tasks tolerating DiffQAS while DiffQAS-dominant tasks degraded when ANO was added.

These findings offer practical guidance: rather than optimizing all circuit components simultaneously, practitioners should diagnose whether the bottleneck is the measurement basis or the gate topology and apply ANO or DiffQAS accordingly---ANO as the safer default on continuous-state tasks since it tolerates DiffQAS, DiffQAS alone when the default topology is mismatched to the reward landscape.

All results are obtained in noiseless state-vector simulation: we do not model finite-shot readout, so the reported expectation values are exact rather than sampled, and shot noise could change the comparisons above. Several directions extend this work. First, evaluating the framework on noisy intermediate-scale quantum (NISQ) hardware would clarify how gate noise, decoherence, and finite-shot readout interact with the ANO and DiffQAS optimizations. Second, matched-output controls (e.g., eight fixed 2-local Hermitians) would help isolate ANO's adaptivity contribution from its locality and output-count changes. Third, expanding the DiffQAS search space to include genuine $R_Z$ and richer entangling primitives, with longer search phases and varied initializations, would better test whether it performs genuine task-specific discovery rather than remaining near initialization. Fourth, complementing BestAvgR with area-under-the-curve summaries and multi-seed confidence intervals would further clarify late-training behavior and the statistical robustness of the differences we report, motivating anti-forgetting mechanisms---such as adaptive target-network schedules or quantum-aware regularization---for the substantial forgetting observed on CartPole.

\section*{\textbf{Acknowledgment}}
{\small
The authors extend their gratitude to the members of the Connectome Lab for their invaluable support and critical contributions to the development of this study. This work was supported by the National Research Foundation of Korea (NRF) grant funded by the Korea government (MSIT) (No. 2021R1C1C1006503, RS-2023-00266787, RS-2023-00265406, RS-2024-00421268, RS-2024-00342301, RS-2024-00435727, RS-2025-25457239, RS-2021-NR061370, NRF-2021M3E5D2A01022515, and NRF-2021S1A3A2A02090597), by the Researchers Program through Seoul National University (No. 200-20250071, 200-20250049, 200-20250116, 200-20250115, 200-20250113, 0670-20250039, 200-20260009). Additional support was provided by the Institute of Information \& Communications Technology Planning \& Evaluation (IITP) grant funded by the Korea government (MSIT) [No. RS-2021-II211343 Artificial Intelligence Graduate School Program, Seoul National University] and by the Global Research Support Program in the Digital Field (RS-2024-00421268). This work was also supported by the Artificial Intelligence Industrial Convergence Cluster Development Project funded by the Ministry of Science and ICT and Gwangju Metropolitan City, by the Korea Brain Research Institute (KBRI) basic research program (25-BR-05-01), by the Korea Health Industry Development Institute (KHIDI) and the Ministry of Health and Welfare, Republic of Korea (HR22C1605), and by the Korea Basic Science Institute (National Research Facilities and Equipment Center) grant funded by the Ministry of Education (RS-2024-00435727). We acknowledge the National Supercomputing Center for providing supercomputing resources and technical support (KSC-2023-CRE-0568, KSC-2024-CRE-0198, KSC-2025-CRE-0340). An award for computer time was provided by the U.S. Department of Energy’s (DOE) ASCR Leadership Computing Challenge (ALCC). This research used resources of the National Energy Research Scientific Computing Center (NERSC), a DOE Office of Science User Facility, under ALCC award m4750-2024, and supporting resources at the Argonne and Oak Ridge Leadership Computing Facilities, U.S. DOE Office of Science user facilities at Argonne National Laboratory and Oak Ridge National Laboratory.
}

\bibliographystyle{ieeetr}
\bibliography{references}

\end{document}